\documentclass{PoS}
\usepackage{booktabs}
\usepackage{adjustbox}
\usepackage{float}
\usepackage{amsmath}
\usepackage{enumitem}
\usepackage{url}
\usepackage{color}
\usepackage[utf8]{inputenc}
\usepackage{tikz}
\usetikzlibrary{shapes.geometric, arrows}
\usepackage[numbers]{natbib}
\title{The fate of axial U(1) in 2+1 flavor QCD towards the chiral limit}

\ShortTitle{The fate of axial U(1) in 2+1 flavor QCD towards the chiral limit}

\author{\speaker{Lukas Mazur}\\
Fakultät für Physik, Universität Bielefeld, D-33615 Bielefeld, Germany\\
E-mail: \email{lmazur@physik.uni-bielefeld.de}}

\author{Olaf Kaczmarek\\
Fakultät für Physik, Universität Bielefeld, D-33615 Bielefeld, Germany\\
E-mail: \email{okacz@physik.uni-bielefeld.de}}

\author{Edwin Laermann\thanks{This work is dedicated to his loving memory}\\
Fakultät für Physik, Universität Bielefeld, D-33615 Bielefeld, Germany\\
}

\author{Sayantan Sharma\\
The Institute of Mathematical Sciences, Chennai 600113, India\\
E-mail: \email{sayantans@imsc.res.in}}

\abstract{The region of the Columbia plot with two light quark flavors is not yet conclusively understood. Non-perturbative effects, 
e.g. the magnitude of the anomalous  U(1) axial symmetry breaking, decides on the nature of the phase transition in this region. We report 
on our study of this region of the Columbia plot using lattice techniques. We use gauge field ensembles generated within the Highly Improved 
Staggered Quark discretization scheme, with the strange quark mass fixed at its physical value 
and the light quark mass varied such that $m_l=m_s/27$ and $m_s/40$, where $m_l=m_s/27$ corresponds to the physical light quark mass. 
We study the eigenvalue spectrum of QCD using the overlap Dirac operator on these gauge field ensembles at finite temperature around the chiral transition temperature $T_c$, as the light quark masses approach the chiral limit, and infer about the fate of the anomalous $U_A(1)$ symmetry breaking.}

\FullConference{The 36th Annual International Symposium on Lattice Field Theory - LATTICE2018\\
22-28 July, 2018\\
Michigan State University, East Lansing, Michigan, USA.}

\begin{document}

\section{The $U_{A}(1)$ puzzle in QCD}
Symmetries determine the order parameters of a phase transition.
The anomalous $U_A(1)$, though not an exact symmetry, is believed to affect the nature of the
phase transition in two flavor QCD~\cite{Pisarski:1983ms}. If $U_A(1)$ is indeed broken 
then the chiral phase transition for $N_f=2$ QCD is expected to be in the $O(4)$ universality class.  
Whether or not $U_A(1)$ is effectively restored at the chiral phase transition can only be
answered non-perturbatively. Lattice studies of $N_f=2$ QCD have tried to address 
this problem. For latest updates see Ref.~\cite{Tomiya:2016jwr} and the talks in this conference~
\cite{lat18jlqcd1, lat18jlqcd2}.

We follow a different approach. Since the up and down quark masses are light compared to the 
intrinsic scale of QCD, the $U_{L}(2)\times U_{R}(2)$ symmetry is only mildly broken. Therefore, 
if we calculate observables that measure the $U_A(1)$ breaking in ($2+1$) flavor QCD with physical 
$u,d$ quark masses near the chiral crossover transition and reduce $m_{u,d}$, we can smoothly reach the 
chiral limit. If indeed $U_A(1)$ is broken, signatures of the $O(4)$ second order line could be observed by
reducing $m_{u,d}$. On the other hand if $U_A(1)$ is effectively restored we should approach the $Z(2)$ 
line or a second order line of $U_L(2)\times U_R(2)/U_V(2)$ universality class~\cite{Pelissetto:2013hqa}. 
In order to verify this we study in detail the eigenvalue spectrum of ($2+1$) QCD with physical $m_{u,d}$ 
and compare the changes to it as the light quark mass is reduced successively. We then study its effects 
on $U_A(1)$ breaking observables as a function of light quark mass near the chiral crossover transition.

\section{QCD Dirac spectrum and $U_{A}(1)$ breaking}
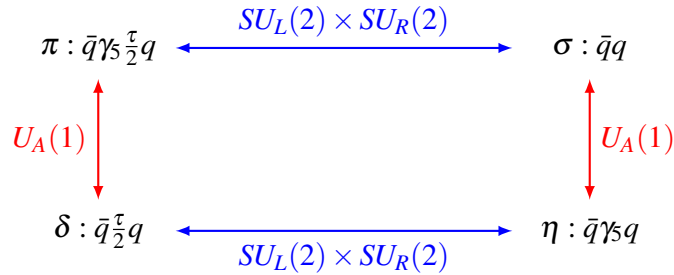
\begin{figure}
\begin{center}
    \tikzstyle{basicNode} = [rectangle, minimum width=5em, minimum height=2em, text ragged, draw=white, fill=white]
\tikzstyle{arrow} = [thick, <->, latex-latex ]

\begin{tikzpicture}[node distance=6em, scale=1.05, transform shape]

    \node(upperLeft)[basicNode]{$\pi:\bar{q}\gamma_{5}\frac{\tau}{2}q$};
    \node(upperRight)[basicNode, right of=upperLeft, xshift=10em]{$\sigma:\bar{q}q$};
    \node(lowerLeft)[basicNode, below of=upperLeft]{$\delta:\bar{q}\frac{\tau}{2}q$};
    \node(lowerRight)[basicNode, right of=lowerLeft, xshift=10em]{$\eta:\bar{q}\gamma_{5}q$};

    \draw [arrow, color=blue] (upperLeft) -- node[anchor=south] { $SU_{L}(2)\times SU_{R}(2)$ } (upperRight);
    \draw [arrow, color=red] (upperLeft) -- node[anchor=east] {$U_{A}(1)$} (lowerLeft);
    \draw [arrow, color=blue] (lowerLeft) -- node[anchor=north] { $ SU_{L}(2)\times SU_{R}(2)$ } (lowerRight);
    \draw [arrow, color=red] (lowerRight) -- node[anchor=west] { $U_{A}(1)$ } (upperRight);

\end{tikzpicture}\hfill
       \vspace{-0.8cm}
        \caption{ Symmetry transformations between scalar and pseudo-scalar mesons. }
       \label{fig:chiPchiD}
    \end{center}
\end{figure}

Unlike chiral symmetry, it is not possible to use expectation values 
of local operators in the case of $U_A(1)$ to investigate its effective breaking or restoration as a function of temperature. Instead, 
it is important to look at the degeneracy of the correlation functions of mesons of appropriate quantum numbers. From Fig. \ref{fig:chiPchiD} 
it is evident that if $U_A(1)$ is effectively restored then the meson correlators along the vertical axes of the box should be degenerate. 
Accordingly it was suggested in Ref.~\cite{Shuryak:1993ee} that the difference of the integrated correlators of pions and 
delta mesons, defined as 
$\chi_{\pi}-\chi_{\delta}=\int d^{4}x\left[\langle i\pi^{+}(x)i\pi^{-}(0)\rangle-\langle \delta^{+}(x)\delta^{-}(0)\rangle\right]$, 
is an observable that will quantify the amount of $U_A(1)$ breaking. This observable contains information 
about the microscopics of QCD since it can be written in terms of the eigenvalue density $\rho(\lambda,m_l)$ with the light quark mass $m_l=m_u=m_d$ of 
the QCD Dirac operator \cite{Hatsuda:1995dy}
\begin{equation}
  \chi_{\pi}-\chi_{\delta}\overset{V\rightarrow\infty}{\longrightarrow}\int_{0}^{\infty}d\lambda\frac{4m_{l}^2\rho(\lambda,m_l)}{(\lambda^{2}+m_{l}^{2})^{2}}.
\end{equation}
In fact, from the study of up to 2-point correlation functions it has been shown~\cite{Bazavov:2012qja,Buchoff:2013nra} 
that when chiral symmetry is restored then the $U_A(1)$ can still be broken if the infrared part of 
eigenvalue density goes as $\lim_{\lambda\rightarrow0}\rho(\lambda,m_l)=\delta(\lambda)m_l^{\alpha}$ with $1<\alpha<2$.
It was further observed \cite{Aoki:2012yj} that if the eigenvalue density is an analytic function in $m$, $\lambda$, then 
using chiral Ward Identities, it is possible to derive many useful properties of the eigenvalue spectrum of QCD at finite 
temperature in the chiral symmetry restored phase. From the study of up to 6-point correlation functions in the 
scalar and pseudo-scalar channels, a sufficient condition for the effective restoration of $U_A(1)$ in $N_f=2$ QCD is that the eigenvalue 
spectrum behaves as $\rho(\lambda,m_l\rightarrow0)\sim \lambda^3$. \\
The properties of the eigenvalue density in QCD in the chiral symmetry restored phase can only be understood from non-perturbative 
lattice studies. 
Specifically, one has to calculate the small eigenvalues of the QCD Dirac operator on the lattice and 
verify that they have near-zero peak as well as parameterize the leading order analytic dependence of the spectrum. 
The existence of a small near-zero peak in the eigenvalue spectrum has been observed and were shown to be strongly sensitive to lattice 
cut-off effects~\cite{Cossu:2015kfa}. Moreover, at high temperatures this near-zero peak arise due to weakly interacting 
instanton, anti-instanton pairs~\cite{Dick:2015twa}. Hence, they are also sensitive to finite volume effects. However, the leading order analytic dependence 
of the infrared spectrum was observed to be quite robust, i.e. insensitive to lattice cut-off effects~\cite{Sharma:2016cmz}. 
In fact, the leading analytic dependence of the eigenvalue spectrum changes from $\rho(\lambda)\sim  \lambda$ 
at chiral crossover transition temperature $T_c$ to $\rho(\lambda)\sim  \lambda^2$ at $1.2~T_c$~\cite{Dick:2015twa} 
for QCD with degenerate $u$-$d$ quark masses slightly heavier than physical values.  In these proceedings, we  
study the eigenvalue spectrum for physical quark masses and its changes as one 
approaches the chiral limit. We also to check if the near-zero peak survives in the chiral limit and how the 
eigenvalue spectrum influences the $U_A(1)$.

\section{Setup}
The gauge field ensembles used in this work were generated within the Highly Improved Staggered Quark (HISQ) discretization scheme  
with 2+1 quark flavors.  The strange quark mass was fixed at its physical value and the light quark mass was varied such that 
$m_l=m_s/27$ and $m_s/40$, where $m_l=m_s/27$ corresponds to the physical quark mass. 
We used gauge field configurations separated by 100 Monte-Carlo trajectories.
The details of the statistics and lattice sizes for the ensembles used for this study 
at each value of light quark mass and temperature are listed in table \ref{tab:setup}. 

\begin{table}[H]
    \vspace{0.2cm}
    \begin{center}
    \begin{tabular}{|c|c|c|c|c|}
    \hline
    $m_s/m_l$	&	  $N_s^3 \times N_{\tau}$	&	  $\beta$	&	 $T/T_c$	&	 \#conf	\tabularnewline
    \hline\hline
	27	&	 $32^3\times 8$	&	 6.390	&	 0.97	&	 69	\tabularnewline
	27	&	 $32^3\times 8$	&	 6.445	&	 1.03	&	 81	\tabularnewline
	27	&	 $32^3\times 8$	&	 6.500	&	 1.09	&	 102	\tabularnewline
	\hline\hline
	40	&	 $32^3\times 8$	&	 6.390	&	 0.99	&	 45	\tabularnewline
	40	&	 $32^3\times 8$	&	 6.423	&	 1.03	&	 50	\tabularnewline
	40	&	 $32^3\times 8$	&	 6.445	&	 1.05	&	 104	\tabularnewline
    \hline
\end{tabular}
\hfill
    \caption{List of the configurations used in this work.}
    \label{tab:setup}
    \end{center}
\end{table}
Staggered fermions suffer from rooting problem and have no index theorem at finite lattice spacings. 
In fact the infrared part of the HISQ eigenvalue spectrum changes quite dramatically with changes in the lattice spacing, showing 
the emergence of a small near-zero peak at the finest lattice spacing ~\cite{Ohno:2012br,Sharma:2018syt}. Instead, we use the 
overlap Dirac operator to measure the low-lying eigenvalue spectrum for the HISQ ensembles. The overlap operator for one massless 
quark flavor is realized as
\begin{equation*}
    D_{ov}  =M\left[1+\gamma_{5}\text{sgn}\left(\gamma_{5}D_{W}(-M)\right)\right],
\end{equation*}
 where $\text{sgn(...)}$ denotes the matrix sign function and $D_W$ is the massless Wilson-Dirac operator with a domain wall 
 height $M\in[0,2)$. Since diagonalizing the overlap operator is computationally expensive, we first estimate the number and 
 the chirality of the zero modes using the gluonic definition of the topological charge $Q=\int d^{4}x\,q(x)$,
with the topological charge density defined as, 
\begin{equation*}
    q(x)=\frac{g^2}{32\pi^{2}}\epsilon_{\mu\nu\rho\sigma}\textrm{tr}\left\{ F_{\mu\nu}(x)F_{\rho\sigma}(x)\right\}.
\end{equation*}
The gluonic definition of the topological charge is valid only for sufficiently smooth configurations.
Thus, we have to apply a smoothing technique on the HISQ configurations to remove the ultra-violet fluctuations 
of the gauge fields before measuring the topological charge. In this work we use Wilson flow \cite{Narayanan:2006rf,Luscher:2010iy} 
to measure $Q$.  The flow smoothens the gauge fields over a region of radius $f=\sqrt{8t}$, $t$ being the flow time. 
For each temperature we adjusted the flow time of the gauge fields such that $fT<1$ and the small instantons 
are not smoothened out.

For some configurations, the number and the chirality of zero modes measured using the Wilson flow was compared to 
the number of zero modes of the overlap Dirac operator. We found agreement in all the cases studied so far ensuring that 
the index theorem is valid. We used the information of the chirality of zero modes measured from the gluonic 
definition to construct a projected overlap operator $D^P_{ov}$ by projecting into the space of eigenvectors having 
the opposite chirality of the zero modes. This ensured that the projected overlap operator had no zero modes and was 
much faster to diagonalize on the lattice. The first $100$ eigenvalues of $D_{ov}^{P\dagger} D_{ov}^P$ on the HISQ 
ensembles were calculated using the Kalkreuter-Simma (KS) Ritz algorithm \cite{Kalkreuter:1995mm}. Finally the eigenvectors of the overlap operator 
were computed using appropriate projections into the full vector-space.

Since the overlap operator was used to measure the eigenvalues of configurations generated with a different fermion 
discretization (HISQ) we renormalized the eigenvalue spectrum and the physical observables appropriately to eliminate 
the effects of this mixed action approach. 
In order to do so we calculated the valence (overlap) strange quark mass assuming that the ratio $m_l/m_s$ is fixed 
for both the valence and the sea sectors. We calculated an appropriate renormalized observable, 
$\Delta=\frac{m_{s}\langle\bar{\Psi}\Psi\rangle_{l}-m_{l}\langle\bar{\Psi}\Psi\rangle_{s}}{T^{4}}$,
for the valence sector using the overlap eigenvalues and independently for the sea quark sector using the traces 
of derivatives and inverse of the HISQ operator using random noise vectors. 
The valence overlap mass $m_s$ was tuned such that these independent estimates of $\Delta$ are equal to each other.

\section{Our Results}
\begin{figure}[]
    \centering
    \begin{minipage}{0.90\textwidth}
        \begin{minipage}{.49\textwidth}
            $m_l=m_s/27$\\
            \raggedleft
            \adjincludegraphics[width=1\linewidth,valign=t]{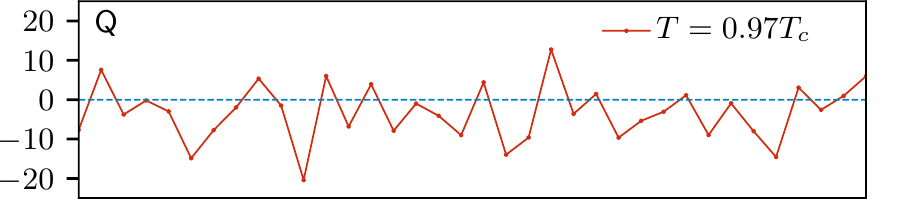}
            \raggedleft
            \adjincludegraphics[width=1\linewidth,valign=t]{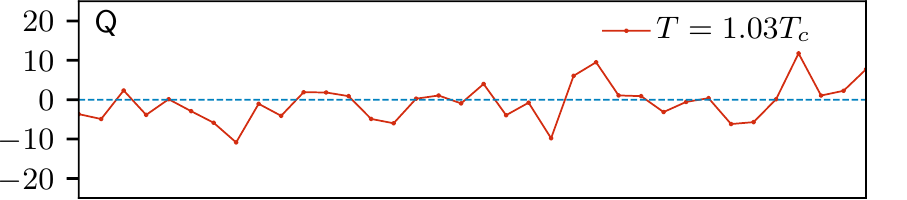}
            \raggedleft
            \adjincludegraphics[width=1\linewidth,valign=t]{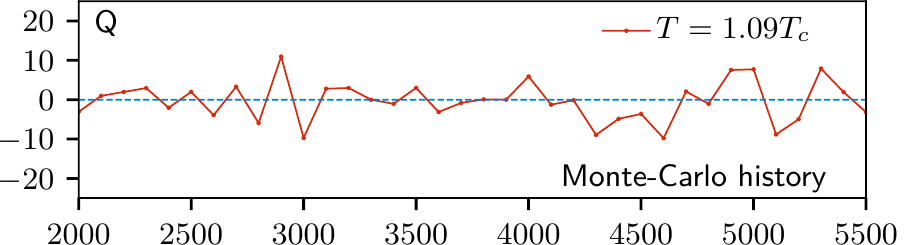}
        \end{minipage}
            \hspace{0.2cm}
            \hspace{0.2cm}
        \begin{minipage}{.49\textwidth}
            $m_l=m_s/40$\\
            \raggedleft
            \adjincludegraphics[width=1\linewidth,valign=t]{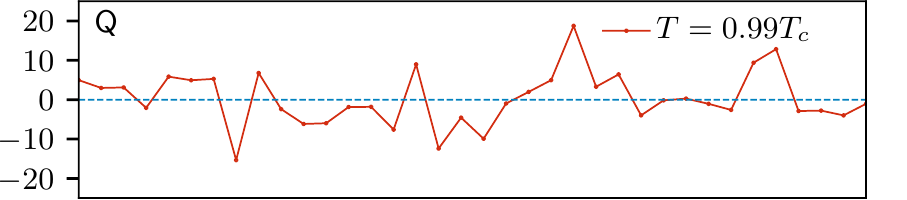}
            \raggedleft
            \adjincludegraphics[width=1\linewidth,valign=t]{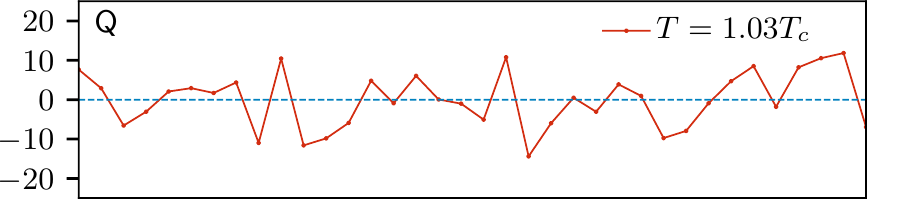}
            \raggedleft
            \adjincludegraphics[width=1\linewidth,valign=t]{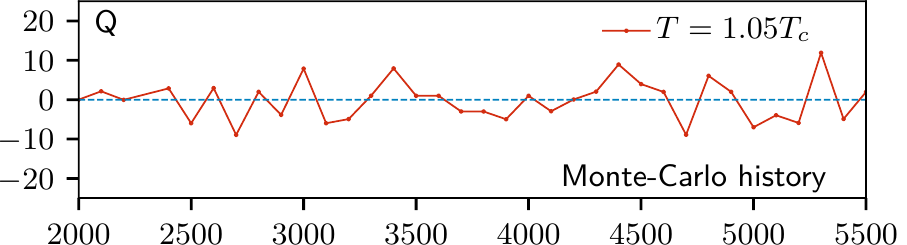}
        \end{minipage}
    \end{minipage}
    \caption{Trajectories of the topological charge of the lattices listed in table \ref{tab:setup}. }
    \label{fig:autocorr}
\end{figure}

\begin{figure}[H]
            \raggedleft
    \adjincludegraphics[width=0.49\linewidth,valign=t]{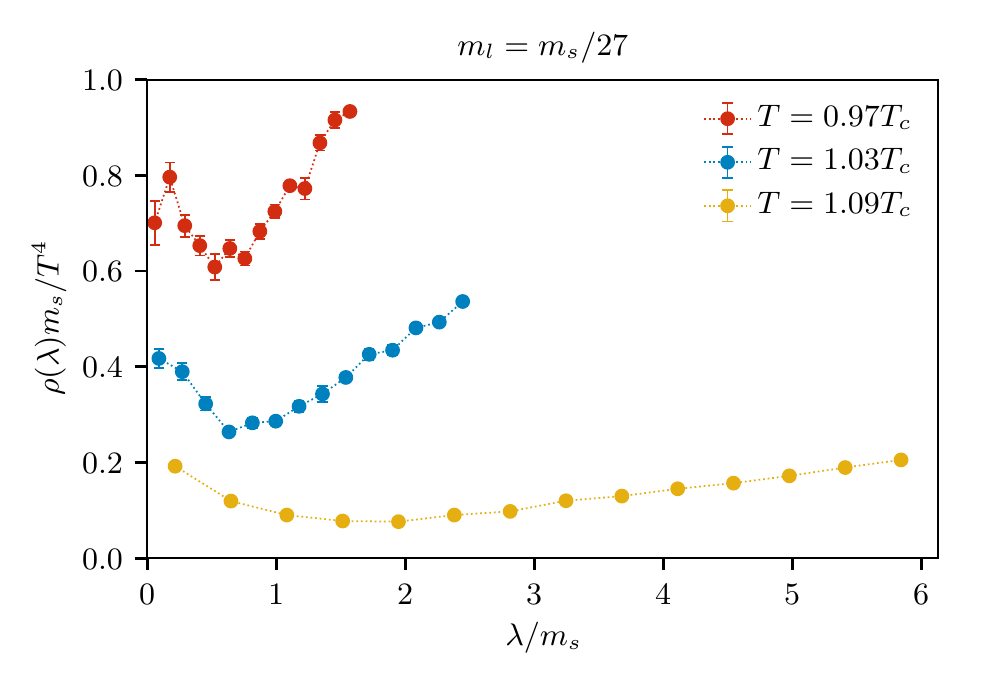}\hfill
            \raggedright
    \adjincludegraphics[width=0.49\linewidth,valign=t]{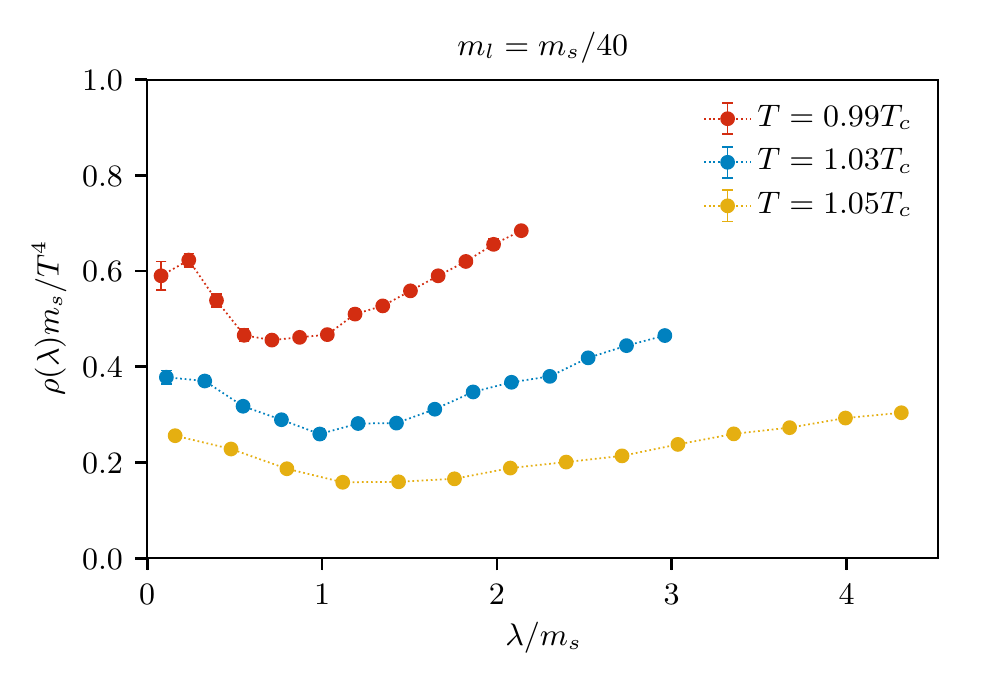}\hfill
    \caption{The eigenvalue density $\rho(\lambda)$ normalized to the tuned valence strange quark mass $m_{s}$.}
    \label{fig:rhoLambda_ms2740}
\end{figure}

As mentioned earlier, we have chosen our configuration at every 100th Monte Carlo step to ensure that our 
results do not suffer from strong autocorrelations. 
To verify this, we plot the trajectory of the topological charge as a function of Monte Carlo time for each 
set of light quark masses and temperatures. Results are summarized in Fig.~\ref{fig:autocorr}. From 
the plots it is evident that we have good sampling of the topological charge in our chosen set of gauge field ensembles.
We measured the eigenvalue spectrum using the overlap operator and renormalized the eigenvalues with the tuned 
valence strange quark mass $m_s$. As discussed earlier this was done such that the resultant density 
of the renormalized eigenvalues $\lambda/m_s$ is not sensitive to the mixed action approach artefacts. Our results 
for the QCD eigenvalue spectrum for two different values of light quark masses are summarized in 
\mbox{Fig. \ref{fig:rhoLambda_ms2740}}. To understand the general features of the spectrum we fit it to the ansatz
 $\frac{\rho(\lambda)}{T^{3}}=\frac{\rho_{0}A}{A^{2}+\lambda^{2}}+c\left|\lambda\right|^{\gamma}$,
where the terms in the R.H.S model the near-zero peak and the leading order analytic dependence of the eigenvalue spectra 
respectively. 
For the physical quark masses, the leading order analytic dependence of the eigenvalue density shown in the left panel of Fig. 
\ref{fig:rhoLambda_ms2740} goes as a power law with the exponent $\gamma\sim1$ within errors for all the 
three temperatures studied. There is a small infrared peak (near-zero peak) whose relative contribution 
is suppressed with increasing temperatures. Lowering the sea light quark mass to $m_s/40$, the 
general features of the eigenvalue spectrum remain unchanged as evident from the right 
panel of Fig. \ref{fig:rhoLambda_ms2740}. The near-zero peak still survives and a linear analytic 
dependence on $\lambda$ is seen as expected from the predictions of chiral perturbation theory. The $m_s/40$ results are 
still preliminary and the finite volume effects are under investigation.

\begin{figure}[H]
    \centering
    \adjincludegraphics[width=0.58\linewidth,valign=t]{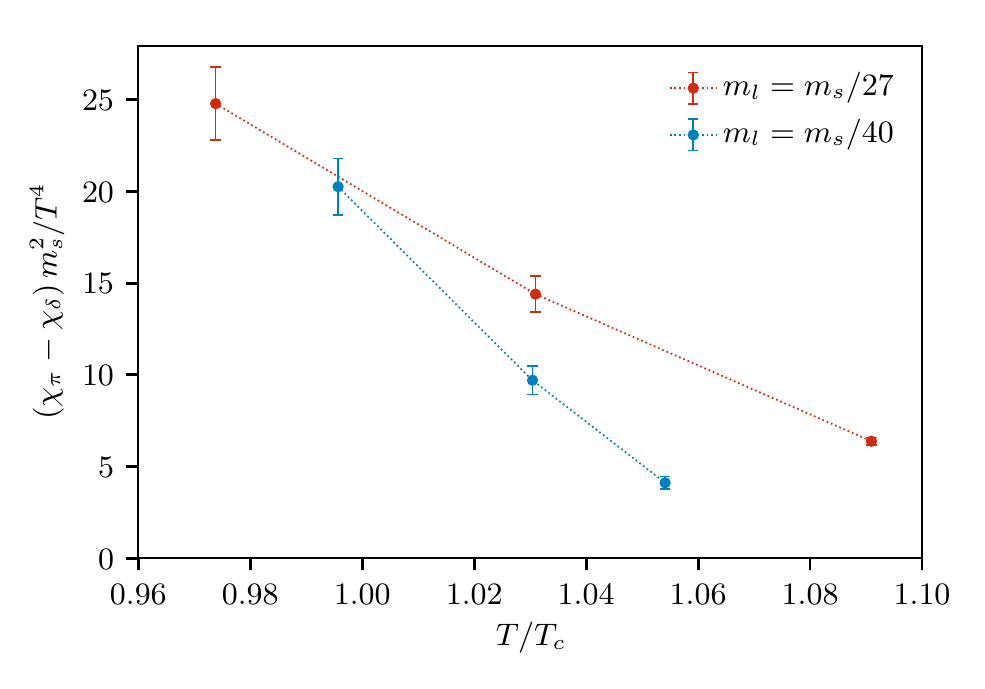}
    \caption{The renormalized $U_{A}(1)$ breaking parameter as a function of quark mass.}
    \label{fig:chiPMinChiDeltaResults}
\end{figure}

In order to study the $U_A(1)$ breaking as a function of temperature we calculated the renormalized 
observable $m_s^2(\chi_{\pi}-\chi_{\delta})/T^4$, in terms of the scaled eigenvalues, which 
are summarized in Fig. \ref{fig:chiPMinChiDeltaResults}. The red and blue points are for lines of 
constant physics corresponding to $m_s/m_l=27$ and $m_s/m_l=40$ respectively. The magnitude of 
$U_A(1)$ breaking at $T\sim T_c$ reduces with the quark mass, but it is still non-zero for $m_l=m_s/40$, 
which in $\overline{\text{MS}}$ scale is $\sim 2$ MeV.
This is in contrast to the results reported in \cite{lat18jlqcd1}, which observes a sudden fall-off 
of this observable to zero for $m_l\lesssim m_l^{phys}$. At present we have only two data points for 
this observable at each temperature, which prevents us from performing a proper chiral extrapolation. 

\section{Conclusions}
In this work we have studied the eigenvalue spectrum of the QCD Dirac operator near $T_c$ with physical value of strange quark mass 
and the light quark mass changed from its physical value $m_s/27$ towards the chiral limit with $m_l=m_s/40$. We do not 
find any drastic change in the features of the eigenvalue spectrum as a function of the light quark mass. The near-zero 
peak at the infrared part of the spectrum survives and the leading analytic dependence of the spectrum is linear in $\lambda$. 
Since we used overlap valence operators to measure the eigenvalue spectrum of the HISQ sea quarks, we appropriately renormalized 
the observables to reduce mixed action effects. We have further observed that the renormalized 
observable $m_s^2(\chi_{\pi}-\chi_{\delta})/T^4$ is still non-zero even for $m_l=m_s/40$ at $T\sim 1.1~T_c$, which shows 
$U_A(1)$ remains broken. We are currently measuring the eigenvalue spectrum and the $U_A(1)$ breaking at 
$m_l=m_s/80$ which will eventually allow us to perform a proper chiral extrapolation and understand the 
$U_A(1)$ puzzle in the chiral limit.

\section{Acknowledgements}
The authors acknowledge support by the Deutsche Forschungsgemeinschaft (DFG) through grant CRC-TR 211, "Strong-interaction matter under extreme conditions".
The computations were performed on the Bielefeld GPU cluster and we acknowledge PRACE for awarding us access to Piz Daint at CSCS, Switzerland.
The GPU codes used in our work were in part based on some publicly available QUDA libraries~\cite{Clark:2009wm}.

\def\bibfont{\footnotesize}

\bibliographystyle{JHEP}
\bibliography{library}
\end{document}